\documentclass{appolb}
\usepackage{epsfig}

\newcommand{\dd}{\mbox{d}}
\newcommand{\reaction}{\mbox{$dp\to\,^{3}\textrm{He}\,\eta$}}

\begin{document}
\title{Measurements on the $^3$He+$\eta$ system at ANKE%
\thanks{Presented at the International Symposium on Mesic Nuclei, Cracow, Poland, 2010}%
}
\author{Alfons Khoukaz
\address{Institut f\"ur Kernphysik, Universit\"at M\"unster, D-48149 M\"unster, Germany }
}
\maketitle
\begin{abstract}
The differential and total cross sections for the
$dp\to\,^{3}\textrm{He}\,\eta$ reaction have been measured in a
high precision high statistics COSY--ANKE experiment near threshold
using a continuous beam energy ramp up to an excess energy $Q$ of
11.3\,MeV with essentially 100\% acceptance. The kinematics
allowed the mean value of $Q$ to be determined to about 9\,keV.
Evidence is found for the effects of higher partial waves for
$Q> 4\,$MeV. The very rapid rise of the total cross section to
its maximum value within $0.5\,$MeV of threshold implies a very
large $\eta\,^3$He scattering length and hence the presence of a
quasi--bound state extremely close to threshold. In addition, differential 
and total cross sections have been measured at excess energies of 
19.5, 39.4, and 59.4~MeV over the full angular range. While at 
19.5~MeV the results can be described in terms of $s$- and $p$-wave 
production, by 59.4~MeV higher partial waves are required. Including 
the 19.5~MeV point together with the near-threshold data in a 
global $s$- and $p$-wave fit gives a poorer overall description of 
the data though the position of the pole in the $\eta^3$He scattering amplitude,
corresponding to the quasi-bound or virtual state, is hardly changed.
\end{abstract}
  
\section{Introduction}

The concept of $\eta$--mesic nuclei was introduced by Liu and
Haider~\cite{Liu}. Since the $\eta$--meson has isospin--zero, the
attraction noted for the $\eta$--nucleon system should add
coherently when the meson is introduced into a nuclear
environment. On the basis of the rather small $\eta$--nucleon
scattering length $a_{\eta N}$ assumed, they estimated that the
lightest nucleus on which the $\eta$ might bind would be $^{12}$C.
Experimental searches for the signals of such effects have
generally proved negative, as for example in the
$^{16}$O$(\pi^+,p)^{15}$O$^*$ reaction~\cite{Chrien}. The larger
$\textit{Re}(a_{\eta N})$ subsequently advocated~\cite{eta-N}
means that the $\eta$ should bind tightly with such heavy nuclei,
generating large and overlapping widths, and thus be hard to
detect~\cite{later}. On the other hand, it also leads to the
possibility of binding even in light systems, such as
$\eta\,^3$He.

Therefore, very precise data on the \reaction\ reaction near
threshold have been taken at the COSY accelerator of
the Forschungszentrum J\"ulich~\cite{Timo,Smyrski,Tobias}. The obtained 
results confirm the energy dependence of the total cross section found in earlier
experiments~\cite{Berger,Mayer}, but with much finer steps in
energy over an extended range. The measurements at the lowest
excess energies $Q$ (the centre--of--mass kinetic energy in the
$\eta\,^3$He system) are of especial interest since they allow to gain 
detailed information about the final state interaction of the meson-nucleus
system.

\section{Experiment}
The experiment was performed with a hydrogen cluster--jet
target~\cite{target} using the ANKE spectrometer~\cite{ANKE}
placed at an internal station of the COoler SYnchrotron
COSY--J{\"u}lich. In case of the measurements very close to threshold 
the deuteron beam energy was ramped slowly and linearly in time,
from an excess energy of $Q = -5.05\,$MeV to $Q = +11.33\,$MeV. 
In addition, for measurements at higher excess energies 
(Q = 19.5, 39.4 and 59.4~MeV)
three fixed values of the beam momentum were requested. 

The $^3$He produced were detected in the ANKE forward detection
system, which consists of two multi-wire proportional chambers,
one drift chamber and three layers of scintillation hodoscopes.
The geometrical acceptance for the $^3$He of interest was $\sim
100$\%, so that systematic uncertainties from acceptance
corrections are negligible. The tracks of charged particles could
be traced back through the precisely known magnetic field to the
known interaction point, leading to a momentum reconstruction for
registered particles. The luminosity required to determine cross
sections was found by simultaneously measuring $dp$ elastic
scattering, with the scattered deuterons being registered in the
forward detector and the proton reconstructed from the missing
mass.

The $^3$He were selected by the $\Delta E/E$ method, with the
$\eta$ meson being subsequently identified through a peak in the
missing--mass distribution~\cite{Ola}. 
In order to determine the differential cross section for each excess
energy, the whole range of the $^{3}\textrm{He}$ c.m. production
angles was divided into individual angular bins and a missing-mass distribution
constructed for each of them. The method to subtract the background below the 
peak of the $\eta$ meson is described in detail in \cite{Timo} and \cite{Tobias}.

\subsection{Results}
\begin{figure}[hbt]
\includegraphics[width=12.6cm]{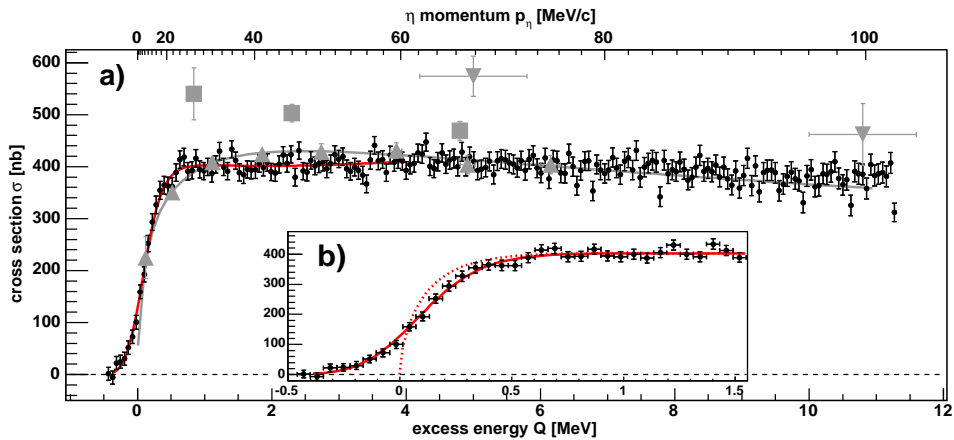}%
\caption{\label{fig:cross} Comparison of the
extracted total cross sections (circles) with previous data drawn
in gray: Ref.~\cite{Berger} (squares), Ref.~\cite{Mayer}
(triangles), and Ref.~\cite{Adam} (inverted triangles). The solid line corresponds to 
a fit to our results for $Q<4\,$MeV considering a strong final state interaction 
as well as the finite COSY beam momentum width. The gray curve is the SPESII fit to their own
data~\cite{Mayer}. Our near--threshold data and fitted curve are
shown in the inset, while the dotted curve is the result to be
expected without the 180\,keV smearing in $Q$.}
\end{figure}

The $dp\to\,^{3}\textrm{He}\,\eta$ total cross sections obtained
at 195 bins in excess energy $Q$ are displayed in
Fig.~\ref{fig:cross}. The minimal relative systematic errors
resulting from the measurement of the excitation function in a
single experiment form a robust data set for any phenomenological
analysis. Our data are broadly compatible with those of
SPESII~\cite{Mayer} and any global difference is within our
overall normalization uncertainty. However, in contrast to our
data presented in Fig.~\ref{fig:cross}b, the SPESII results do not
define firmly the energy dependence in the near--threshold region.
The total cross section reaches its maximum value within 0.5\,MeV
of threshold and hardly decreases after that. This behavior is in
complete contrast to phase--space expectations and indicates a
very strong final state interaction~\cite{Wilkin}.

\begin{figure}[hbt]
\begin{center}
\includegraphics[width=8cm]{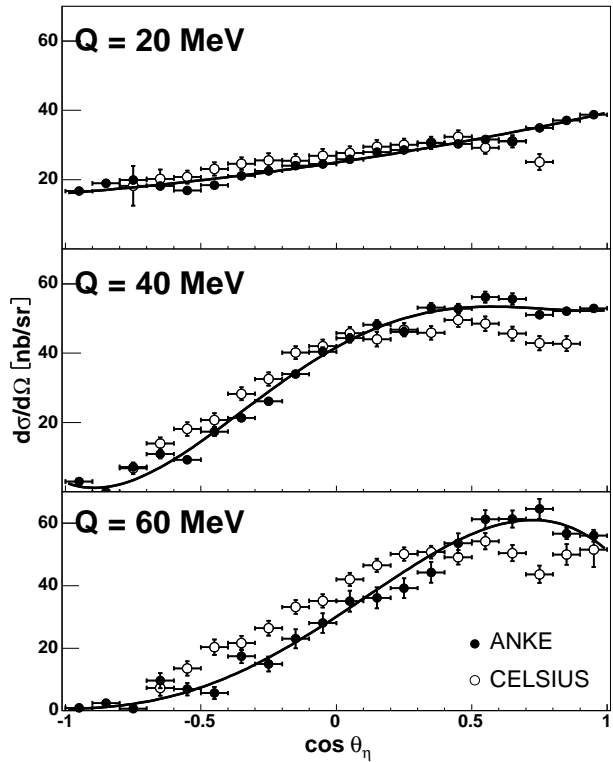}
\caption{Differential cross sections for the three excess energies
studied at ANKE (filled circles). The CELSIUS data (open circles)
shown in the 60~MeV plot were measured at 80~MeV~\cite{Bil02}. The
solid lines represent polynomial fits to the ANKE
data.\label{dcs}}
\end{center}
\end{figure}

Figure~\ref{dcs} shows the angular distributions obtained at the
three highest energies. Also presented are the points measured in a
missing-mass experiment by a CELSIUS collaboration in the vicinity of
$Q=20$~MeV and 40~MeV, as well as those at 80~MeV~\cite{Bil02}. These
are generally in agreement with the present results, though our data
have smaller statistical error bars and cover the complete
$\cos\theta_{\eta}$ range. There is no sign of a forward dip at
19.5~MeV and that any at 39.4~MeV is much weaker than the one found
at CELSIUS. However, although the statistics were poorer and the
number of points fewer, the CELSIUS group also measured events where
the $\eta$ meson was detected through its two photon decay in
coincidence with the $^3$He. These data also seemed to show less of a
forward dip at both 20 and 40~MeV~\cite{Bil02}.
Figure~\ref{dcs} also shows polynomial fits to our points.

\begin{figure}[hbt]
\begin{center}
\includegraphics[width=12.6cm]{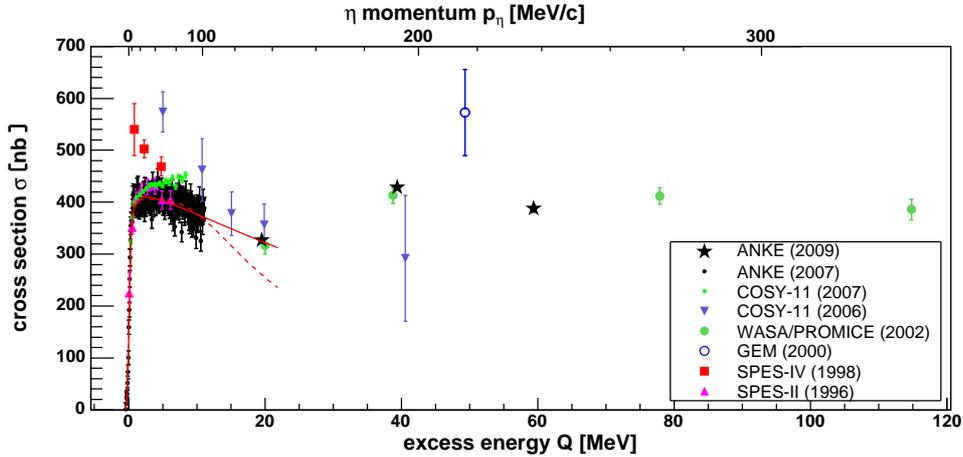}
\caption{Total cross section for the
$dp\to\,^{3}\textrm{He}\,\eta$  reaction. The results from this
experiment (black stars) have an additional overall 15\% systematic
uncertainty that is largely common with our previous
data~\cite{Timo} (small black circles). Also shown are data from
Ref.~\cite{Smyrski}, Ref.~\cite{Bil02}, Ref.~\cite{Berger},
Ref.~\cite{Mayer}, Ref.~\cite{Bet00}, Ref.~\cite{Adam}. The
solid and dashed lines show the result of the recursive fit to the
data with and without considering the 19.5~MeV point. \label{sigt}}
\end{center}
\end{figure}

With the very precise ANKE data the excitation function near threshold
is now given by Figure~\ref{sigt}.  

In order to prove that a nearby pole in the complex $Q$ plane is
responsible for the unusual energy dependence of the
\reaction\ cross section, it is necessary to show that the pole
induces a change in the phase as well as in the magnitude of the
$s$--wave amplitude. Since the cross section is proportional to
the absolute square of the amplitude, much phase information is
thereby lost. However, as will be shown 
the interference between the $s$-- and $p$--waves,
as seen in the newly published angular
distributions~\cite{Timo,Smyrski}, leads to the required
confirmation.

\begin{figure}
\begin{center}
\includegraphics[width=8cm]{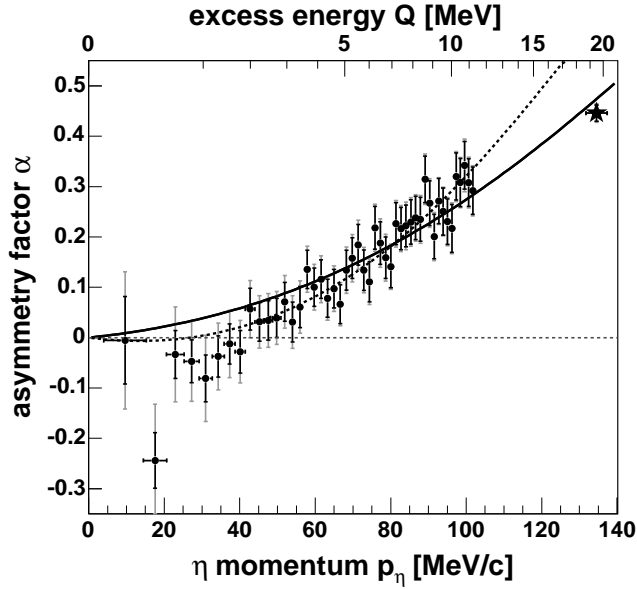}
\caption{Slope parameter $\alpha$ of Eq.~(\ref{alpha}) as a function
of the $\eta$ c.m.\ momentum. Bin widths and statistical errors are
shown bold, systematic uncertainties with feint lines. The solid and
dashed lines show the result of the recursive fit to the data with
and without considering the data point at 19.5~MeV.\label{angas}}
\end{center}
\end{figure}

The \reaction\ differential cross sections close to threshold 
were found to be linear
in $\cos\theta_{\eta}$, where $\theta_{\eta}$ is the c.m.\ angle
between the initial proton and final $\eta$. Throughout the range
of the new COSY measurements, $Q< 11$\,MeV~\cite{Timo,Smyrski},
there is no sign of the $\cos^2\theta_{\eta}$ term that is needed
for the description of the angular distributions at higher
energies~\cite{Bil02}. The angular dependence may therefore be
summarised in terms of an asymmetry parameter $\alpha$, defined as
\begin{equation}
\label{alpha}
\alpha=\left.\frac{\dd\phantom{x}}{\dd(\cos\theta_{\eta})}
\ln\left(\frac{\dd\sigma}{\dd\Omega}\right)\right|_{\cos\theta_{\eta}=0}\:.
\end{equation}
The variation of the ANKE measurements of $\alpha$ with the $\eta$
momentum $p_{\eta}$ is shown in Fig.~\ref{angas}.

On kinematic grounds, the angular dependence near threshold might
be expected to develop like $\vec{p}_p\cdot\vec{p}_{\eta}=
p_pp_{\eta}\cos\theta_{\eta}$, where $\vec{p}_p$ and
$\vec{p}_{\eta}$ are the c.m.\ momenta of the incident proton and
final $\eta$--meson, respectively. However, one striking feature
of Fig.~\ref{angas} is that, although $\alpha$ rises sharply with
$p_{\eta}$, it only does so from about 40\,MeV/c instead of from
the origin, as one might expect on the basis of the above
kinematic argument. At low values of $p_{\eta}$ the error bars are
necessarily large and $\alpha$ might even to go negative in this region.
This feature is not inconsistent with the results of other
measurements~\cite{Smyrski,Mayer} that have different systematic uncertainties 
and so it is possibly a genuine effect. Part of this non--linear behaviour
arises from the steep decrease in the magnitude of the $s$--wave amplitude
with momentum. However, the size of the effect observed can only arise
through the rapid variation of the phase of this amplitude, of the
type generated by a nearby pole in the complex $Q$ plane.

There are two independent \reaction\ $s$--wave amplitudes ($A$ and
$B$)~\cite{GW} and five $p$--wave though, to discuss the data
phenomenologically, we retain only the two ($C$ and $D$) that give
a pure $\cos\theta_{\eta}$ dependence in the cross section. The
production operator
\begin{equation}
\label{amps} \hat{f}=A\,{\vec{\varepsilon}}\cdot\hat{p}_p
+iB\,(\vec{\varepsilon}\times\vec{\sigma})\cdot\hat{p}_p
+C\,\vec{\varepsilon}\cdot\vec{p}_{\eta}
+iD\,(\vec{\varepsilon}\times\vec{\sigma})\cdot\vec{p}_{\eta}\,
\end{equation}
has to be sandwiched between $^3$He and proton spinors. Here
$\vec{\varepsilon}$ is the polarisation vector of the deuteron.
The corresponding unpolarised differential cross section depends
upon the spin-averaged value of $|f|^2$
\begin{equation}
\label{cross1} \frac{\dd\sigma}{\dd\Omega}=
\frac{p_{\eta}}{p_p}\,\overline{|f|^2}=\frac{p_{\eta}}{3p_p}\,I\,.
\end{equation}
Using the amplitudes of Eq.~(\ref{amps}) this yields
\begin{equation}
I=|A|^2+2|B|^2+p_{\eta}^{\:2}|C|^2 +2p_{\eta}^{\:2}|D|^2+
2p_{\eta}Re(A^*C+2B^*D)\cos\theta_{\eta}, \label{dsdo2}
\end{equation}
which has the desired linear dependence on $\cos\theta_{\eta}$,
with an asymmetry parameter
\begin{equation}
\alpha=2p_{\eta}\,\frac{Re(A^*C+2B^*D)}
{|A|^2+2|B|^2+p_{\eta}^{\:2}|C|^2+2p_{\eta}^{\:2}|D|^2}\:\cdot
\label{alpha2}
\end{equation}

The strong $\eta^3$He final--state interaction that gives rise to
the quasi--bound pole should affect the two $s$--wave amplitudes
$A$ and $B$ in a similar way and some evidence for this is to be
found from the deuteron tensor analysing power $t_{20}$, 
which is small and seems to change little, if at all, from near
threshold to their measurement limit 16.6\,MeV above
threshold~\cite{Berger}. As a consequence, $|A|\propto |B|$ throughout our
range of interest and it is plausible to represent the data using a
spin--average amplitude. However, further experiments to measure both 
the deuteron tensor analysing powers and spin--correlations will be undertaken at
COSY--ANKE~\cite{Tobias2,SPIN}.

In the original fit to the whole of the ANKE \reaction\ total
cross section data~\cite{Timo} shown in Fig.~\ref{fig:cross}, any
influence of $p$--waves was neglected and the data represented by
\begin{equation}
f_s = \frac{f_B}{(1-p_\eta/p_1)(1-p_\eta/p_2)}\,, \label{eq:fsi2}
\end{equation}
with
\begin{eqnarray}
\label{mom}
\nonumber
f_B &=& (50\pm8)\,(\textrm{nb/sr})^{1/2}\,,\\
\nonumber
p_1 &=& [(-5\pm7^{+2}_{-1})\pm i(19\pm2\pm1)]\,\textrm{MeV/c}\,,\\
p_2 &=& [(106\pm5)\pm i(76\pm13^{+1}_{-2})]\,\textrm{MeV/c}\,.
\end{eqnarray}
The first error bar is statistical and the second, where given,
systematic. The error on $f_B$ is dominated by the 15\% luminosity
uncertainty~\cite{Timo}. Note that only the first pole (at
$p_\eta=p_1$) is of physical significance and for this unitarity
requires that $\textit{Re}(p_1)<0$. The signs of the imaginary
parts of the pole positions are not defined by the data. As will
be seen later, the position of the first pole remains stable when
fitting simultaneously the angular dependence and the total cross
section. In contrast, the second pole is introduced to parametrise
the residual energy dependence, which can arise from the reaction
mechanism as well as from a final--state interaction.

Equation~(\ref{eq:fsi2}) shows an $s$--wave amplitude whose phase
and magnitude vary quickly with $p_{\eta}$, but we expect that,
apart from the momentum factor, the $p$--wave amplitudes should be
fairly constant. In the absence of detailed analysing power
information, we take $A=B=f_s$ and $C=D$ to be a complex constant.
With these assumptions the total cross section and
asymmetry parameter become \cite{Wilkin}:%
\begin{eqnarray}
\nonumber \sigma&=&\frac{4\pi
p_{\eta}}{p_p}\left[|f_s|^2+p_{\eta}^{\,2}|C|^2\right],\\
\alpha&=&2p_{\eta}\,\frac{Re(f_s^*C)}
{|f_s|^2+p_{\eta}^{\:2}|C|^2}\:\cdot \label{alpha3}
\end{eqnarray}
If the phase variation of the $s$--wave amplitude is neglected, by
replacing $f_s$ by $|f_s|$, the best fit to the asymmetry
parameter considering data up to Q = 11 MeV
does display a little curvature due to the falling of
$|f_s|^2$ with $p_\eta$. Nevertheless, it fails badly to reproduce shape of the
low--momentum data.

On the other hand, when the phase variation of $f_s$ given by
Eq.~(\ref{eq:fsi2}) is retained, the much better description of
the data given by the dashed line in Fig.~\ref{angas} is achieved,
with no degradation in the description of the total cross section
presented in Fig.~\ref{fig:cross}. Furthermore, the difference in
the behaviour of $\alpha$ in the low and not--so--low momentum
regions can now be easily understood. The parameters of the fit
are
\begin{eqnarray}
\label{mom2} \nonumber
f_B &=& (50\pm8)\,(\textrm{nb/sr})^{1/2}\,,\\
\nonumber
C/f_B &=& [(-0.47\pm 0.08\pm0.20) + i(0.33\pm 0.02\pm0.12)]\,(\textrm{GeV/c})^{-1}\,,\\
\nonumber
p_1 &=& [(-4\pm7^{+2}_{-1})- i(19\pm2\pm1)]\,\textrm{MeV/c}\,,\\
p_2 &=& [(103\pm4) -i(74\pm12^{+1}_{-2})]\,\textrm{MeV/c}\,.
\end{eqnarray}
An inclusion of the 19.5 MeV point leads to the solid line of 
Fig.~\ref{angas} and Fig.~\ref{sigt}, which gives
a much poorer overall description of the near-threshold data though the
position of the pole in the $\eta^3$He scattering amplitude,
corresponding to the quasi-bound or virtual state, is hardly changed.

\begin{figure}
\includegraphics[width=12.6cm]{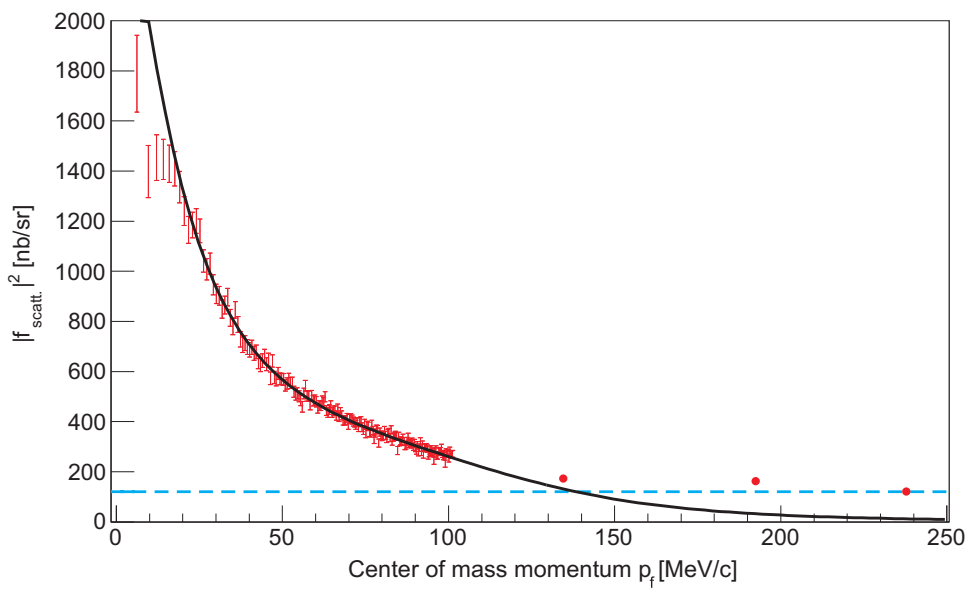}
\caption{Scattering amplitude squared extracted from the data obtained at ANKE 
as function of the $\eta$ c.m.\ momentum. 
The solid line shows the result of a fit to the near-threshold data.\label{slope}}
\end{figure}

Since the
overall phase is unmeasurable, it is permissable to take the $f_B$
of Eq.~(\ref{eq:fsi2}) to be real. Furthermore, because of the
interference between the $s$-- and $p$--waves, the relative phases
of $C$, $p_1$, and $p_2$ do now influence the observables, though
the differential cross section remains unchanged if the signs of
all the imaginary parts are reversed. Compared to the original
solution, where the effects of the $p$--waves were
neglected~\cite{Timo}, the position of the nearby pole $p_1$ is
little changed. This is hardly surprising because this parameter
is mainly fixed by the data from a region which is dominated by
the $s$--waves. Less expected is the very modest change in the
position of $p_2$, which could have been affected more by the
introduction of $C$. As a consequence, the $\eta\,^3$He scattering
length is also changed only marginally to $a=(\pm
10.9+1.0\,i)$\,fm, where the two signs of $\textit{Re}(a)$ again
reflect the possibility of either a quasi--bound or a virtual
state. These results nicely explain the strong decrease of the scattering 
amplitude squared extracted from the data obtained at ANKE 
as function of the $\eta$ c.m.\ momentum (Fig.~\ref{slope}).

The ANKE data indicate that the $s$--wave amplitude for \reaction\
undergoes a very rapid change of phase in the near--threshold
region of the type expected from the presence of a quasi--bound or
virtual $\eta\,^3$He state. 

\begin{thebibliography}{99}
%
\bibitem{Liu} Q.\,Haider and L.C.\,Liu, Phys.\ Rev.\ C \textbf{34} (1986)
1845.
%
\bibitem{Chrien} R.E.\,Chrien et al., Phys.\ Rev.\ Lett.\ \textbf{60} (1988)
2595.
%
\bibitem{eta-N} M.\,Batini\'c et al., Phys.\ Rev.\ C \textbf{51} (1995)
2310 [arXiv:nucl-th/9501011], \emph{idem} arXiv:nucl-th/9703023.
%
\bibitem{later}
See, for example: C.\,Garcia-Recio, J.\,Nieves, T.\,Inoue, and
E.\,Oset, Phys.\ Lett.\ B \textbf{550} (2002) 47
[arXiv:nucl-th/0206024]; M.\,Post, S.\,Leupold, and U.\,Mosel,
Nucl.\ Phys.\ A \textbf{741} (2004) 81 [arXiv:nucl-th/0309085].
%
\bibitem{Timo} T.\,Mersmann et al., Phys.\ Rev.\ Lett.\
\textbf{98} (2007) 242301.
%
\bibitem{Smyrski} J.\,Smyrski \textit{et al.}, Acta Physica Slovaca
\textbf{56}, 213 (2006).
%
\bibitem{Tobias} T.\,Rausmann et al., Phys.\ Rev.\ C
\textbf{80} (2009) 017001.
%
\bibitem{Berger} J.\,Berger et al., Phys.\ Rev.\ Lett.\ \textbf{61}
(1988) 919.
\bibitem{Mayer} B.\,Mayer et al., Phys.\ Rev.\ C \textbf{53} (1996)
2068.
%
\bibitem{target}
A.\,Khoukaz \textit{et al.}, Eur.\ Phys.\ J.\ D \textbf{5} (1999) 275.
%
\bibitem{ANKE}
S.\,Barsov \textit{et al.}, Nucl.\ Instr.\ Meth.\ A \textbf{462} (2001)
364.
%
\bibitem{Ola}  A.\,Wro\'{n}ska \textit{et al.}, Eur.\ Phys.\ J.\
A \textbf{26} (2005) 421.
%
\bibitem{Wilkin} C.~Wilkin \textit{et al.}, Phys.\ Lett.\ B \textbf{654} (2007) 92.
%
\bibitem{Adam} H.-H.\,Adam \textit{et al.}, Phys.\ Rev.\ C
\textbf{75}, 014004 (2007).
%
\bibitem{Bil02} R.~Bilger \textit{et al.}, Phys.\ Rev.\ C
\textbf{65} (2002) 044608.
%
\bibitem{Bet00} M.\,Betigeri \textit{et al.}, Phys.\ Lett.\ B
\textbf{472} (2000) 267.
%
\bibitem{GW} J.-F.\,Germond and C.\,Wilkin, J.\ Phys.\ G \textbf{14}
(1988) 181.
%
\bibitem{Tobias2} T.\,Rausmann \& A.\,Khoukaz, COSY proposal \#157
(2006);\\
www.fz-juelich.de/ikp/anke/en/proposals.shtml.
%
\bibitem{SPIN} A.\,Kacharava, F.\,Rathmann, and C.\,Wilkin (ANKE
  Collaboration), COSY proposal \#152, nucl-ex/0511028.
%

\end{thebibliography}
\end{document}